# EXPLORING STUDENT'S BLENDED LEARNING THROUGH SOCIAL MEDIA

**Leon Andretti Abdillah**

Information System Department, Faculty of Computer Science, Bina Darma University,
Jln. Ahmad Yani No.3, Plaju, Palembang, 30264
leon.abdillah@yahoo.com

## ABSTRACT

*Information technology (IT) has been used widely in many aspects of our daily life. Social media as a leading application on the internet has changed many aspects of life become more globalized. This article discussed the use of social media to support learning activities for students in the faculty of computer science. The author used Facebook and WordPress as an alternative to electronic learning, those were: 1) online attendance tool, 2) media storage and dissemination of course materials, 3) and event scheduling for the lectures. Social media succeed to change the way of modern learning styles and environment. The results of this study are some learning activities such as (1) Preparation, (2) Weekly meeting activities, (3) Course Page, (4) Social Media as Online Attendance Tool, (5) Social Media as Learning Repository and Dissemination, and (6) Social Media as Online Event Scheduling. Change conventional learning model becomes visual and distanceless.*

## INTRODUCTION

Social media (SM) offers many opportunities in various sectors. Especially in globalization era the requirement for the development of information technology (IT) makes the world of education must adopt and involve IT in the learning process (Abdillah, 2013). In the previous research, the author used e-learning software (Moodle) and blog (WordPress) for student learning environment.

There is some doubts whether or not Facebook can be used to support educational activities (Roblyer *et al.*, 2010). A study has suggested that in the future there should be concerns about Facebook integration with education and teaching (Bicen & Cavus, 2011). In this research, the author expands the media for learning purposes by involving social media, Facebook. Because these internet sites attract billions of cyber users, especially young adults (SocialBakers, 2014).

Table 1 Facebook User Distribution Based on SocialBakers

| No | Country | Facebook user based on age | | Facebook user based on gender | |
|---|---|---|---|---|---|
| | | **The Largest** | **The Second Largest** | **Male** | **Female** |
| 1. | USA | 25-34 | 18-24 | 46% | 54% |
| 2. | India | 18-24 | 25-34 | 76% | 24% |
| 3. | Brazil | 18-24 | 25-34 | 47% | 53% |
| 4. | **Indonesia** | **18-24** | **25-34** | **59%** | **41%** |
| 5. | Mexico | 18-24 | 25-34 | 50% | 50% |

The current condition of student learning behaviors is affected by the development of new social technology, such as social media. Based on Table 1, it is shown that the greatest number of Facebook users are young people aged between 18 to 24 years. This age range is an age when they are studying in college. Earlier observation that the author caught in a number of lectures showed the activity of social media Facebook usage while being in the classroom. See this condition; the authors are keen to develop the learning environment by involving social technologies that are popular in this decade.

The goal of this article is to manage student learning environment via social media. This article discusses the alternative media for blended e-learning strategy. The alternative media is using social media and blogs. This combination will create and support "always on" learning environment.

Based on the data collected and observations of the activities, the authors will discuss several strategies to conduct blended learning. Blended learning is the thoughtful integration of classroom face-to-face learning with online learning (Garrison & Kanuka, 2004). Those integrations could include the mix or combination of (1) modes of web-based technology, (2) various pedagogical approaches, (3) any form of instructional technology, and (4) instructional technology with actual job tasks to create a harmonious effect of learning and working (Driscoll, 2002). Today, organizations have a myriad of learning approaches and choices (Singh, 2003), see Table 2. The blended concept of learning is associated with: (1) thinking less about delivering instruction and more about producing learning, (2) reaching out to students through distance education technologies, and (3) promoting a strong sense of community among learners (Rovai & Jordan, 2004).

Table 2 Learning approaches and choices

| Learning approaches | Aproaches |
|---|---|
| Synchronous physical formats | • Instructor-led Classrooms & Lectures<br>• Hands-on Labs & Workshops<br>• Field Trips |
| Synchronous online formats (live e-learning) | • Online Meetings<br>• Virtual Classrooms<br>• Web Seminars and Broadcasts<br>• Coaching<br>• Instant Messaging<br>• Conference Calls |
| Self-paced, asynchronous formats | • Documents & Web Pages<br>• Web/Computer Based Training Modules<br>• Assessments/Tests & Surveys<br>• Simulations<br>• Job Aids & Electronic Performance Support Systems (EPSS)<br>• Recorded Live Events<br>• Online Learning Communities and Discussion Forums<br>• Distributed and Mobile Learning |

The author will cluster the discussions into three groups, preparations, weekly activities meeting, classical meeting, and students participations via social media (Facebook) and blog (Wordpress).

As of April 2014, Indonesia is the world's fourth position in terms of the number of Facebook users (Abdillah, 2014) after USA, India, and Brazil. Young adults (18-24) people dominate Facebook users in Indonesia followed by the users in the age of 25-34 (Table 1)**.** In Indonesia, most Facebook users are young people aged between 18 to 24 years. The age range is an age periods of the teens in college. There are 59% male users and 41% female users in Indonesia, compared to 46% and 54% in the USA, 76% and 24% in India, 47% and 53% in Brazil, and 50% and 50% in Mexico.

Social media like facebook and/or twitter become one of the most visited and used applications over the internet. They become one of the top internet application recently (Bi *et al.*, 2014). The rapid development of online social networks has tremendously changed the way of people communicating with each other (Rahadi & Abdillah, 2013). They offer borderless environment for various activities. They change how people communicating each other, promoting their products (Baumgartner & Morris, 2010), becoming conventional communications venues for young adults (Abdillah, 2014), campaigning political parties (Abdillah, 2014; Bi *et al.*, 2014; SocialBakers, 2014), disseminating knowledge and information (Abdillah, 2014), etc. This article will discuss the use of social media in managing online learning environment for higher education students in computer science field.

There is strong evidence that social media can facilitate the creation of Personal Learning Environment (PLE) (Dabbagh & Kitsantas, 2012) that helps learners to aggregate and share the results of learning achievements, participate in collective knowledge generation, and manage their meaning making. Several authors have found that the most efficient teaching model is a blended approach, which combines self-paced learning, live e-learning, and face-to-face classroom learning (Alonso, López, Manrique, & Viñes, 2005). By combining classical face-to-face meeting with online social media, learning process become placeless, borderless and timeless.

The combination of these schemes provides collaborative learning environment for students. Classical meetings are very important to get in touch directly with the students. Facebooking is the activity where the learning atmosphere involves cyber atmosphere for both lecturers and students. Lecturers are able to disseminate the knowledge (learning materials), assignments, online attendance to the students. On the other hand, students are also able to participate in learning scenario, downloading course materials, write their online presence in every meeting, submit their assignments, give comments, or discuss with their team.

Another technology used in this study is known as "social software." A blog, one of social software, is often involved in education domain (Kim, 2008). A study has summarized that students use blogs as an effective aid in teaching and learning (Williams & Jacobs, 2004). Because, blogs are easily published and accessed via the internet (Beldarrain, 2006), easy to use for everyone (Sanjaya & Pramsane, 2008), as practical tool for higher education (Al-Fadda & Al-Yahya, 2010), and last but not least allowing students to learn interactively and collaboratively (Schroeder *et al.*, 2010). In this research, the author uses Wordpress as one of the most popular blogging tool or weblog software. WordPress is a free installation resource which has many useful plug-ins, comment spam-fighting features, and user-friendly interface (Hong, 2008), open-source and is relatively easy to use (Bean, 2014). WordPress is also very popular content management system that has been deployed in educational settings especially taking advantage of its potential with Multisite environments (Hoover, 2015). Lecturers use WordPress to publish course materials to students. Students create a blog post that a weekly question related to a main topic covered that week (Quesenberry *et al.*, 2014).

Thus, this article focuses on how to optimize and to use current popular social technology application (Facebook and Wordpress) in conducting learning activities in higher education in Indonesia. The rest of this article will be organized in three sections or discussions related to (1) methods (section two), (2) discussions (section three), and (3) conclusions (section four), then list of the references used in this study.

# METHODS

This research involves a total 100 students participated in the observations. The author made observations of student behavior in the classroom for some time previously. Furthermore, the authors explore a number of features on social media Facebook who could be involved in the learning process.

The subject is one specialization subject in computer science/information systems field, Knowledge Management Systems (KMS) (Abdillah, 2014). In KMS subject there is a subject of Knowledge Sharing. The author uses Social Media to apply knowledge sharing between lecturer and students, and student with their group study in the class. Knowledge sharing process (de Jorge Moreno, 2012) is the key to enhance the externalization and dissemination of knowledge.

Social media (Facebook), weblog application (Wordpress), mobile operating systems for Apple (iOS), blog application for iPhone (Wordpress for iPhone), and Facebook application for iPhone as the main software are used in the research. The author also involves cloud storage, Dropbox. Cloud storage services like Dropbox (Meske *et al.*, 2014) allows users to store files remotely and to synchronize them with multiple devices. DropBox makes a class become a learning community working together simultaneously, in real-time, with or without face-to-face contact (Ries *et al.*, 2012).

Lecturer combines severals conditions to conduct the blended learning involved class meeting, e-learning, and group presentations. The total general meeting is 14 (fourteen) meetings. Two of them are e-learning meetings; two meetings use for exams (9 and 14), one meeting for group presentations (13). Every class meeting will involve 31 – 55 students. During the semester, a subject consist of six main activities, namely: Daily Attendances, Daily Test (5), Mid Test (9), Weekly Reports (2-12), Presentation (13), and Final Test (14).

# RESULTS AND DISCUSSIONS

In this section, Author explains how to set-up social media-based learning environments. The discussions are made in 6 sections as follows: (1) Preparation, (2) Weekly meeting activities, (3) Course Page, (4) Social Media as Online Attendance Tool, (5) Social Media as Learning Repository and Dissemination, and (6) Social Media as Online Event Scheduling.

In the first stage or preparation phase, lecturer needs to install an iOS App for Facebook and Wordpress (Figure 1), to prepare blog address, to create a Facebook page for "Knowledge Management Systems", and to set up the array of serial meetings through one semester (approx. 3 – 4 months or 14 – 16 meetings). The course material for each meeting will be supplied before the meeting in the most popular article format, pdf (Abdillah, 2012) and/or ppt forms. Students need to download and read every single material carefully. Lecturer uploads the course materials through Facebook page and blog. The number of power point slides for each week of the meeting at least 18 (eighteen) and at most 37 slides.

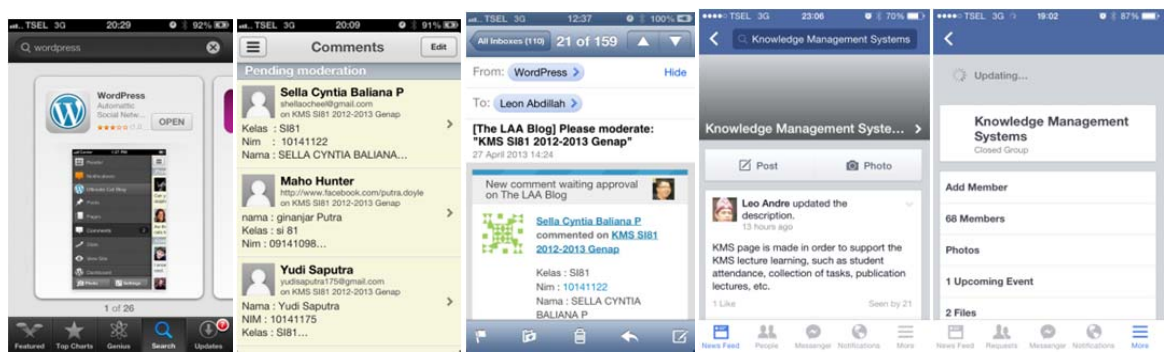

Figure 1 iOS App for Wordpress and Facebook

In the second stage for weekly activity, lecturer needs to manage the weekly schedule. In the first meeting, lecturer will normally introduce himself, materials of the course, Facebook and blog address. Lecturer also needs to explain course rules, review for the pre-requested subjects (optional), and introducing the current materials, grade scale systems, references. At the end of the first meeting, lecturer needs to set group study for the class. For small class, every group consists of 2 – 4 students. For medium class, every group consists of 4 – 6 students. For big class, every group consists of 6 – 8 students. After finishing setting up the group in the classroom, then the next step is to determine the topic of each group.

For the first month, learning activities will be dominantly classical face-to-face meeting. While giving the knowledge in the oral traditional class meeting, this activity also involves setting up group discussions in every class, asking students to create or activate or opening blog and Facebook accounts. Lecturer supplies course materials via blog and facebook. In the second month, lecturer no longer focuses on the classical meeting, instead of more active in Facebook facilities, such as
(1) online attendance, (2) checking submitted assignments, (3) chat with online participant students, (4) involve students activities by using blogs, among others. In the third month, the author starts using social media for students to prepare their presentation and assignments submissions, and also to publish the post related to their team project reports or works, as seen in Figure 2.

The author creates a Facebook page to mediate the communication between lecturer and students. The author also creates an e-mail for the course. The author chooses to set the facebook page as close page. It means if anyone would like to join the facebook page, they need approval from the admin. Facebook page enriches with some features, such as (1) Write Post, (2) Add Photo or Video, (3) Ask Questions, and (4) Add File. Facebook page also provide some tabs of Members, Events, Photos, and Files. After the course Facebook page had created, then lecturer asked students to join the page. In order to keep the page only for the students in the class, lecturer needs to set the page as "Closed Group" page. This setting will grant only invited, or registered students access to the cyber class. Moreover, lecturers are able to accept their requests. If needed, the lecturer could supply the page with some descriptions and tags for the page. When setting up a Facebook page, lecturer could grant the post for every member or only for admin. In this study only admin could post some information into course page.

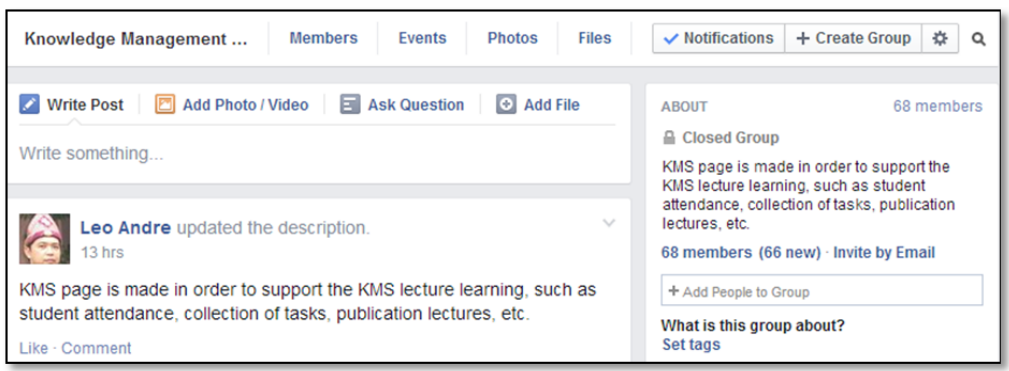

Figure 2 Facebook Page for the Course of Knowledge Management Systems

The author uses Facebook page to record students attendance in the class and for students who are not in the class. Facebook page provides a feature of "asked a question." The author uses this feature to share a question where the students need to reply in the comment box to write their class code, name, and student number. For students that both attend in the class and Facebook will get point 1, otherwise 0.5. To navigate the information related to the online attendance, the author informs the meeting, date, material theme, and notes. This strategy is very useful to engage students who can't join the class physically, so they are still able to join the class and receive the course materials. Students in the class are excited to participate in cyber class. Figure 3 shows the example of some students who inform their attendance via social media group.

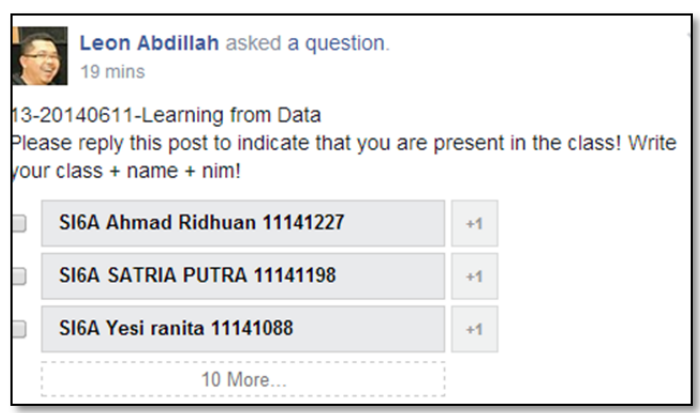

Figure 3 Online Attendance Check List

Course materials could also be stored in social media. The feature of "Add File" enables the author to upload and store the course materials. The uploaded and stored fields automatically notify all of the course participants (see Figure 4). Facebook offers 2 (two) kinds of document upload. The first method is by using the button of "Create Doc." This method helps the author to create an online document via Facebook. Another method is by using the button of "Upload File." This method helps the author to upload a file by using a dialog box. In this study, the author uses the second facility by uploading the course materials (in pdf format). Lecturer names each course material with the label of "Lecture_1" until "Lecture_12" followed by lecture title, and so on. These facilities create an online repository for students and helped lecturers to disseminate their course materials. Beside using Facebook, the author also uses Weblog (blog) to disseminating the knowledge through the internet (Abdillah, 2014). Figure 4 shows some uploaded or stored course materials into virtual social media class.

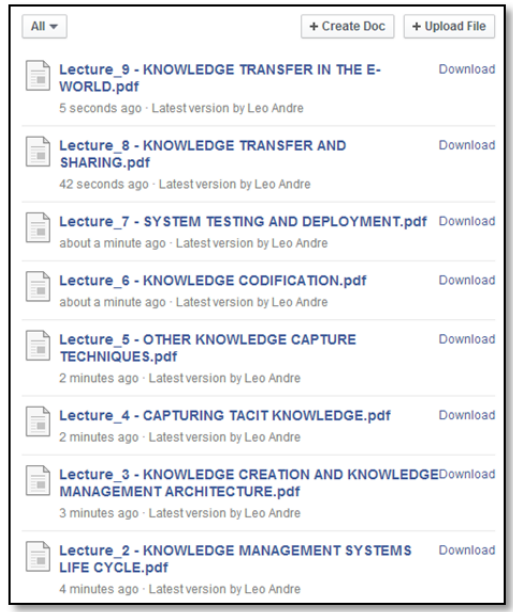

Figure 4 Uploaded or Stored Files for Knowledge Management Systems

In every semester, a course consists of some activities including (1) students' attendance, (2) daily questionnaires, (3) mid-semester exam, (4) teamwork students project presentation, (5) final semester exam, and others important announcements. Normally, at the beginning of the semester, there is tentative schedule for one semester. However, sometimes, the schedule could be changed. In order to accommodate this condition, the author uses social media to create events related to any announcements to the students. Once an event has been set, all of the participants will be notified. After the invitations have been sent to the participants, students as participants may respond to several conditions, such as: going, maybe, or invited (see figure 5).

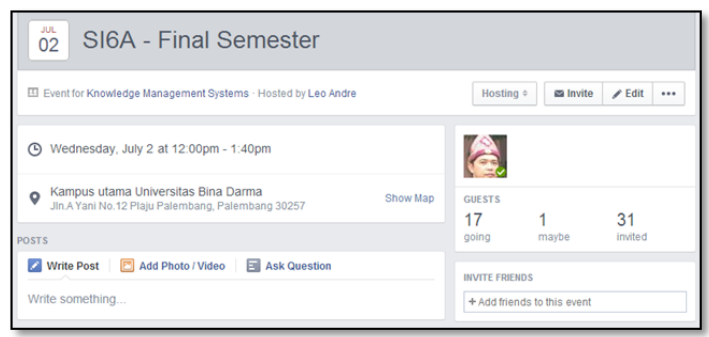

Figure 5 Online Event Scheduling for Knowledge Management Systems

## CONCLUSIONS

Social media as one of Internet application offers many benefits to modern learning systems. It has changed the way of learning styles and approaches. Social media could be used to support IT or cyber-based university. The author extends IT based learning environment by adopting the power social media besides e-Learning and blogs.

Based on the experiences and observations, the author suggests some schemes by using social media: (1) Social media as online attendance tool, (2) Social media as learning repository and dissemination, and (3) Social media as online event scheduling. Social media changes conventional learning model become visual and distanceless. Lecturers are able to disseminate course materials to learners (students), managing online attendance, explore the activities of each the participants such as assignment submissions, etc. In this article, the author also involves iPhone smartphone to monitor the student's activities on Facebook and blog.

For the next research, the author is still interested in working the online learning environment but with some extensions involving the benefits various of widgets and mobile devices. The next research involves more applications and enriches the learning strategy and environment. More features of distant learning with mobile based atmosphere are also to be considered for the next study. Last but not least, the author is interested in integrating IT based education with psychology aspects.